\documentclass[%
 showpacs,
 amsmath,amssymb,
]{revtex4-1}

\usepackage{graphicx}
\usepackage{dcolumn}
\usepackage{bm}

\usepackage{CJK}

\usepackage{color}
\usepackage{amssymb}
\usepackage{amsfonts}

\usepackage[colorlinks,urlcolor=blue,linkcolor=blue,citecolor=blue,anchorcolor=blue]{hyperref}

\begin{document}

\preprint{APS/123-QED}

\title{Unveiling the link between fractional Schr\"odinger equation and light propagation in honeycomb lattice}

\author{Da Zhang$^1$}
\author{Yiqi Zhang$^{1,2}$}
\email{zhangyiqi@mail.xjtu.edu.cn}
\author{Zhaoyang Zhang$^{1,2}$}
\author{Noor Ahmed$^{1}$}
\author{Yanpeng Zhang$^{1}$}
\author{Fuli Li$^2$}
\author{Milivoj R. Beli\'c$^3$}
\author{Min Xiao$^{4,5}$}
\affiliation{%
$^1$Key Laboratory for Physical Electronics and Devices of the Ministry of Education \& Shaanxi Key Lab of Information Photonic Technique,
Xi'an Jiaotong University, Xi'an 710049, China \\
$^2$Department of Applied Physics, School of Science, Xi'an Jiaotong University, Xi'an 710049, China\\
$^3$Science Program, Texas A\&M University at Qatar, P.O. Box 23874 Doha, Qatar \\
$^4$Department of Physics, University of Arkansas, Fayetteville, Arkansas 72701, USA \\
$^5$National Laboratory of Solid State Microstructures and School of Physics, Nanjing University, Nanjing 210093, China
}%

\date{\today}

\begin{abstract}
\noindent
  We establish a link between the fractional Schr\"odinger equation (FSE) and light propagation in the honeycomb lattice (HCL) ---
the Dirac-Weyl equation (DWE).
The fractional Laplacian in FSE causes a modulation of the dispersion relation of the system,
which in the limiting case becomes linear.
In the HCL, the dispersion relation is already linear around the Dirac point,
suggesting a possible connection with the FSE.
Here, we demonstrate this connection by describing light propagation in both FSE and HCL, using DWE.
Thus, we propagate Gaussian beams according to FSE, HCL around the Dirac point, and DWE, to discover very similar behavior ---
the conical diffraction.
However, if an additional potential is brought into the system, the link between FSE and HCL is broken,
because the added potential serves as a perturbation,
which breaks the translational periodicity of HCL and destroys Dirac cones in the dispersion relation.
\end{abstract}

\pacs{03.65.Sq, 42.70.Qs}
\maketitle


The fractional Schr\"odinger equation (FSE) is the fundamental equation of the fractional quantum mechanics \cite{laskin.pla.268.298.2000,laskin.pre.62.3135.2000,laskin.pre.66.056108.2002}.
As compared to the standard Schr\"odinger equation, it contains the fractional Laplacian operator instead of the usual one.
This change brings profound differences in the behavior of the wave function.
In optics, the fractional Laplacian corresponds to a non-parabolic dispersion,
which means that the dispersion of the system is directly modulated.
Interesting phenomena based on the FSE were reported in the past few years \cite{zhang.prl.115.180403.2015,zhang.sr.6.23645.2016,zhang.lpr.10.526.2016,liemert.math.4.31.2016},
and some related nonlinear aspects were also discussed \cite{guo.jmp.53.083702.2012,klein.prsa.470.20140364.2014,zhang.oe.24.14406.2016}.
The complicated fractional Laplacian operation in the FSE is made more manageable
if one uses the Fourier transform method in both theory and experiment \cite{longhi.ol.36.2883.2015};
however, the real problem is the lack of real physical systems described directly by the FSE.
Here, we aim at pointing to such a system.

On the other hand, the topological photonics \cite{lu.np.8.821.2014} --- as a new field ---
has experienced an explosive development and still attracts great attention.
Among different photonic models that are explored, the honeycomb lattice (HCL, viz. the photonic graphene) \cite{rechtsman.nature.496.196.2013,plotnik.nm.13.57.2014}
has excited a particular interest.
Research on HCL has inspired new ideas to develop new techniques and methods in optical manipulation, image transmission,
and optical trapping, to name a few.
The goal of this paper is to establish a link between the FSE and the HCL,
which seemingly are not related.
The inspiration for this investigation comes from the fact that conical diffraction can be observed in both FSE and HCL.
Therefore, one has reasons to believe that the cause behind might be similar in the two systems.
Indeed, the dispersion around the Dirac point in HCL is nearly linear \cite{neto.rmp.81.109.2009},
which indicates that the dispersion is effectively modulated ---
a consequence that can also arise in FSE, due to the fractional Laplacian.

In this paper, we first demonstrate that the FSE
can be transformed into the Dirac-Weyl-like equation,
and then construct the HCL by using the three-wave interference method of light propagation.
The corresponding band structure is calculated by using the plane-wave expansion method.
Next, we numerically simulate light propagation in the FSE,
the Dirac-Weyl equation (DWE) and the HCL, and note apparent similarities that point to similar origins.
Two typical cases --- direct and oblique excitation of the Bloch modes of the Dirac cone ---
are discussed in some detail.
Finally, we give a discussion on the breakup of the model when harmonic potential is added,
leading to the symmetry breaking and the disappearance of the Dirac point in the band structure.
We believe that our research may pave a way in the exploration of real physical systems
that can be described by the FSE directly.


We start with the two-dimensional FSE without potential \cite{zhang.prl.115.180403.2015,zhang.sr.6.23645.2016,zhang.lpr.10.526.2016}
\begin{equation}\label{eq1}
i\frac{{\partial {\psi}}}{{\partial z}} - \left(-\frac{\partial^2}{\partial x^2}-\frac{\partial^2}{\partial y^2}\right)^{\alpha/2} \psi=0,
\end{equation}
for the slowly-varying envelope $\psi$ of the optical field.
Here, $z$ is the normalized propagation distance, and $x$ and $y$ are the scaled transverse coordinates;
finally, $\alpha$ is the L\'{e}vy index ($1<\alpha\leq2$).
When $\alpha=2$, one recovers the usual Schr\"{o}dinger equation in free space.
We will consider the opposite limiting case $\alpha=1$, as the most interesting one \cite{longhi.ol.36.2883.2015}.
We assume that the field envelope $\psi$ can be written in the component form
\[
\psi=
\begin{bmatrix}
\psi_+\\
\psi_-
\end{bmatrix},
\]
which comes from the possible factorization of the Laplacian.
Namely, if one writes the usual Laplacian operator as
\begin{equation}\label{eq2}
\hat{\mathcal{L}}=-\left(\frac{\partial^2}{\partial x^2}+\frac{\partial^2}{\partial y^2}\right)I,
\end{equation}
where $I$ is the $2\times2$ unit matrix,
one can factorize it as
\begin{equation}\label{eq3}
\hat{\mathcal{L}}=\left( \hat{\beta} \hat{\mathcal{L}}_+ \right) \left( \hat{\beta} \hat{\mathcal{L}}_- \right),
\end{equation}
where
$
\hat{\mathcal{L}}_+=\partial_x+i\partial_y,
$
$
\hat{\mathcal{L}}_-=-\partial_x+i\partial_y,
$
and $\hat{\beta}$ is a $2\times2$ Hermite matrix composed of constant elements.
Plugging Eqs. (\ref{eq2}) and (\ref{eq3}) into Eq. (\ref{eq1}),
one obtains
\begin{equation}\label{eq4}
  \left(i\frac{\partial }{\partial z}\right)^2 = \left( \hat{\beta} \hat{\mathcal{L}}_+ \right) \left( \hat{\beta} \hat{\mathcal{L}}_- \right),
\end{equation}
which can be formally rewritten as
\begin{equation}\label{eq5}
  \left. i\frac{\partial}{\partial z} \right/ \hat{\beta} {\hat{\mathcal{L}}_+} = \hat{\beta} {\hat{\mathcal{L}}_-} \left/ i\frac{\partial}{\partial z} \right..
\end{equation}
It is reasonable to assume that both sides of Eq. (\ref{eq5}) can be made equal to a non-zero constant $c$.
For convenience, we assume that $c$ is a positive constant.
As a result, one ends up with
\begin{align}\label{eq6}
i\frac{\partial }{\partial z} \psi =
\begin{bmatrix}
0 & c \mathcal{L}_+ \\
\mathcal{L}_-/c & 0
\end{bmatrix}
\psi.
\end{align}
We note here that the matrix $\beta$ is adopted as the Pauli matrix $\sigma_x$ in the derivative of Eq. (\ref{eq6}).
Clearly, Eq. (\ref{eq6}) is a Dirac-Weyl-like equation,
which describes a fermion with fractional spin that is determined by $c$.
If $c=1$, the DWE (\ref{eq6}) describes the spin-1/2 fermions.
In fact, the value of $c$ is indeed 1, because the speed of the spread of the two components should be the same,
which leads to $|\mathbf{k}|/c=c|\mathbf{k}|\Rightarrow c=1$.

It is interesting to note that the DWE can be obtained from the usual Schr\"odinger equation
with a potential described by the HCL at the Dirac points \cite{ablowitz.pra.79.053830.2009,ablowitz.pra.82.013840.2010,song.nc.6.6272.2015,song.2dm.2.034007.2015}.
The propagation of light in such a HCL can be described by the usual Schr\"odinger equation \cite{ablowitz.pra.79.053830.2009,ablowitz.siam.72.240.2012}
\begin{equation}\label{eq7}
i \frac{\partial \psi}{\partial z} + \nabla^2\psi + V_h(x,y)\psi = 0,
\end{equation}
in which the Laplacian is $\nabla^2=\partial^2_x+\partial^2_y$ and
$V_h(x,y)$ is a periodical potential that can be connected with the intensity pattern of the three interfering plane waves.
In other words, the missing link between Eqs. (\ref{eq1}) and (\ref{eq7}) is the DWE (\ref{eq6});
i.e., the propagation dynamics according to the FSE with $\alpha=1$ can be
mimicked by the Schr\"odinger equation around Dirac points in the HCL.

\begin{figure}[htpb]
\centering
\includegraphics[width=0.45\columnwidth]{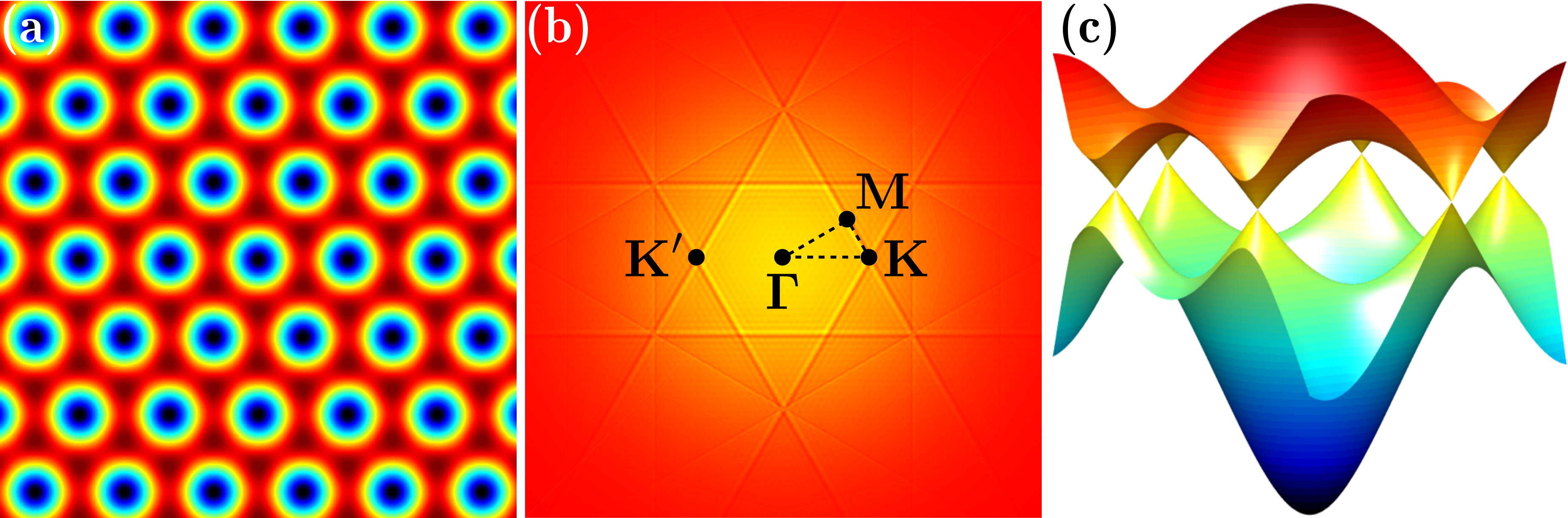}
\caption{(a) Honeycomb lattice resulting from the three-wave interference with $k_0=1$.
(b) Brillouin zone spectroscopy due to the far-field diffraction pattern,
in which the high-symmetry points $\mathbf \Gamma$, \textbf{M}, $\mathbf{K}$ and $\mathbf{K}'$ of the first Brillouin zone are displayed.
(c) The corresponding band structure with $V_0=1$.}
\label{fig1}
\end{figure}

The periodic potential that results from the
intensity pattern of the three interfering plane waves \cite{boguslawski.pra.84.013832.2011} can be written as
\begin{equation}\label{eq8}
V_h(x,y) = V_0 \left(9-\left| \sum_{j=1}^3 \exp \left(ik_0{\bf b}_j \cdot {\bf r}\right) \right|^2\right),
\end{equation}
where $V_0$ indicates the input beam intensity,
$k_0$ is used to adjust the lattice constant, and ${{\bf b}_1 = (1,\,0) }$,
${{\bf b}_2 = ( { -{1}/{2},\,\sqrt{3}/{2}} )}$, and ${{\bf b}_3 = ( { - {1}/{2},\, -{{\sqrt 3 }}/{2}} ) }$
are the three unit vectors used to build the lattice.
The HCL obtained from Eq. (\ref{eq8}) is displayed in Fig. \ref{fig1}(a).
One can calculate the lattice constant --- the distance between the two sites --- to be $4\pi/(3\sqrt{3}k_0)$.
According to the far-field diffraction patterns \cite{bartal.prl.94.163902.2005,freedman.nature.440.1166.2006,liu.apb.99.727.2010,zhang.pra.81.041801.2010},
one can obtain the corresponding Brillouin zone spectroscopy of the HCL as shown in Fig. \ref{fig1}(b),
in which the high symmetry points of interest, $\mathbf \Gamma(0,0)$, $\mathbf{M}(3k_0/4,\sqrt{3}k_0/4)$, $\textbf{K}(k_0,0)$ and $\mathbf{K}'(-k_0,0)$ are separately labeled.

To construct the corresponding band structure, one can adopt the plane-wave expansion method \cite{peleg.prl.98.103901.2007,ablowitz.pra.79.053830.2009,zhang.lpr.9.331.2015,zhang.lpr.10.526.2016}.
The solution of Eq. (\ref{eq7}) can be written as $\phi_n(\mathbf{r};\mathbf{k})\exp[i\beta_n(\mathbf{k})z]$,
in which $\phi_n(\mathbf{r};\mathbf{k})$ is the Bloch mode and $\beta_n(\mathbf{k})$ is the corresponding propagation constant.
Plugging this solution into Eq. (\ref{eq7}), one obtains
\begin{equation}\label{eq9}
  -\beta_n \phi_n + \nabla^2\phi_n + V_h(x,y)\phi_n = 0,
\end{equation}
which is an eigenvalue problem.
The calculated band structure is displayed in Fig. \ref{fig1}(c).
As expected, there are six Dirac cones at the high-symmetry points $\mathbf{K}$ and $\mathbf{K}'$ in the first Brillouin zone.

\begin{figure}[htpb]
\centering
\includegraphics[width=0.45\columnwidth]{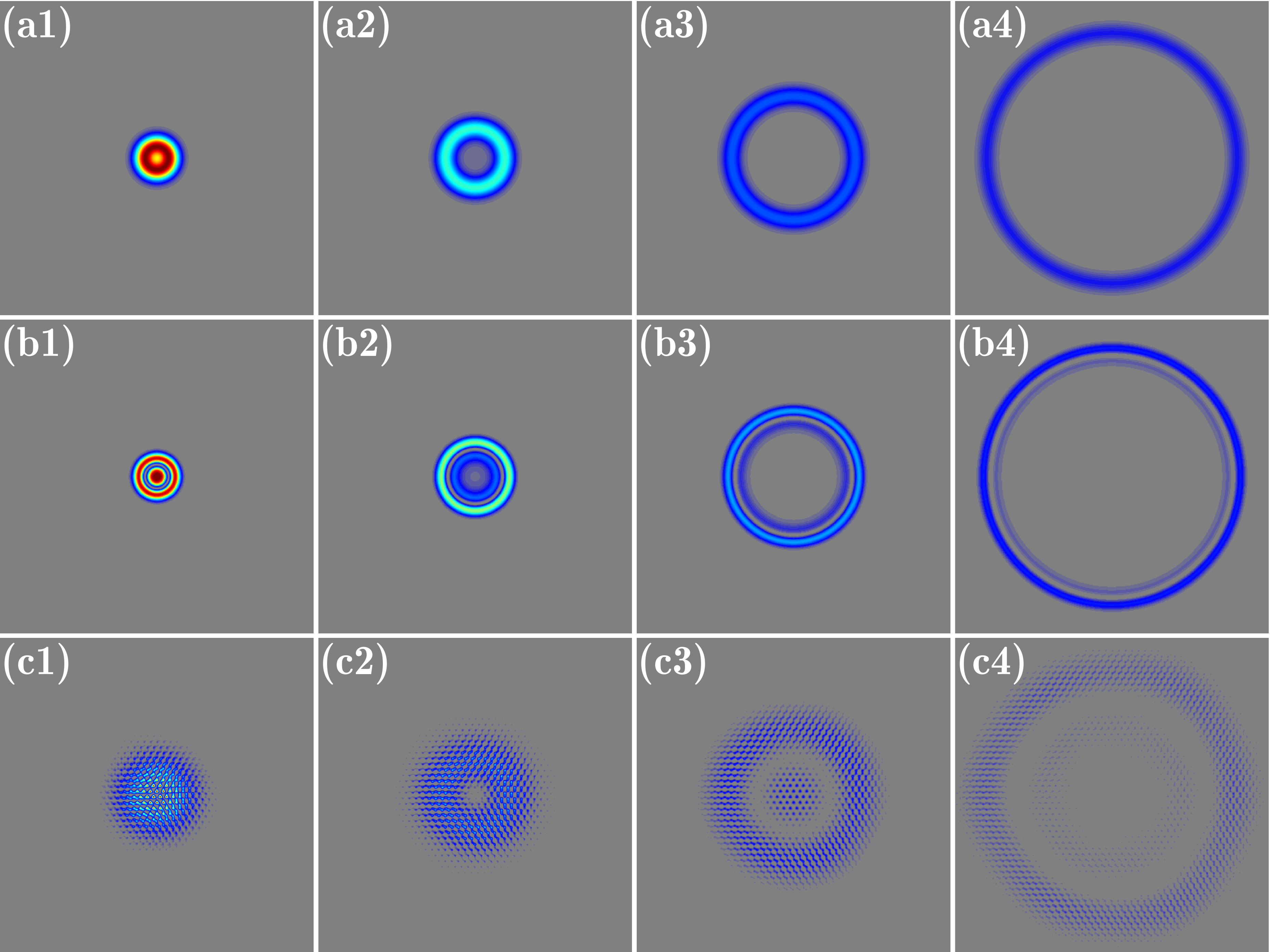}
\caption{(a1)-(a4) Intensity distribution of light propagating according to FSE at $z=10$, 20, 40 and 80.
(b1)-(b4) Same as (a1)-(a4), but according to DWE.
(c1)-(c4) Same as (a1)-(a4), but according to HCL.
The scale dimension of all panels is $100\times100$. }
\label{fig2}
\end{figure}

In the following, we will numerically demonstrate that light propagation in the FSE
can be well mimicked by the propagation in the HCL.
In addition, the inadequacy of such mimicking is also discussed,
once an additional potential is included.
We first consider and compare the propagation of Gaussian beams in the FSE, DWE, and HCL.
This is displayed in Fig. \ref{fig2}.

The propagation of a Gaussian beam $\psi_0=\exp(-r^2/25)$ according to FSE is presented in Figs. \ref{fig2}(a1)-\ref{fig2}(a4).
Since the dispersion relation is linear,
the light undergoes a conical diffraction, as reported previously \cite{zhang.sr.6.23645.2016}.
As a comparison, we also display the intensity distributions
according to DWE (\ref{eq6}) and to HCL (\ref{eq7}),
in Figs. \ref{fig2}(b1)-\ref{fig2}(b4) and Figs. \ref{fig2}(c1)-\ref{fig2}(c4), respectively.
We would like to note that we only excite the component $\psi_+$ by a Gaussian beam $\psi_0=\exp(-r^2/25)$,
and in Figs. \ref{fig2}(b1)-\ref{fig2}(b4) only the component $\psi_+$ is shown.
To excite the mode of the Dirac cone of the HCL and obtain the conical diffraction in Figs. \ref{fig2}(c1)-\ref{fig2}(c4),
we launch the three beam interference pattern multiplied by a wide Gaussian beam $\psi_0=\exp(-r^2/400)$ into one site of the HCL \cite{song.nc.6.6272.2015}, since there are two sites in one unit cell.
As expected, the conical diffraction is observed in all three cases.
The appearance of wider and less resolved rings in Figs. \ref{fig2}(c1)-\ref{fig2}(c4) is caused by the use of a wider Gaussian beam.
Still, similar behavior is observed.

It is interesting to point out that the spreading speeds of the three conical diffractions in Fig. \ref{fig2} are almost the same.
For the first two cases, one can find that the relation between the radius of the ring $r$ and the propagation distance $z$ is $r/z=1$,
if one performs the Fourier transform of Eqs. (\ref{eq1}) and (\ref{eq6}).
As concerns the third case, the spreading speed is different from that of the discrete model;
in the continuum model, the spreading speed will be controlled by the potential coefficient $V_0$ (which here equals 1).

Therefore, according to numerical simulations,
the link between the FSE and the HCL indeed exists --- it is the DWE.
In other words, the HCL represents potentially a real physical system that can be described by the FSE.
Thus far, such real physical systems have been absent from the literature.
Still, the observable difference between the two is the existence of a Poggendorff's dark ring \cite{berry.po.50.13.2007}
in the conical diffraction according to the HCL.
The reason is that there are two cones with opposite chiralities at the Dirac point in HCL,
while there is only one cone in the FSE (as shown in Fig. \ref{fig4} below).

\begin{figure}[htpb]
\centering
\includegraphics[width=0.45\columnwidth]{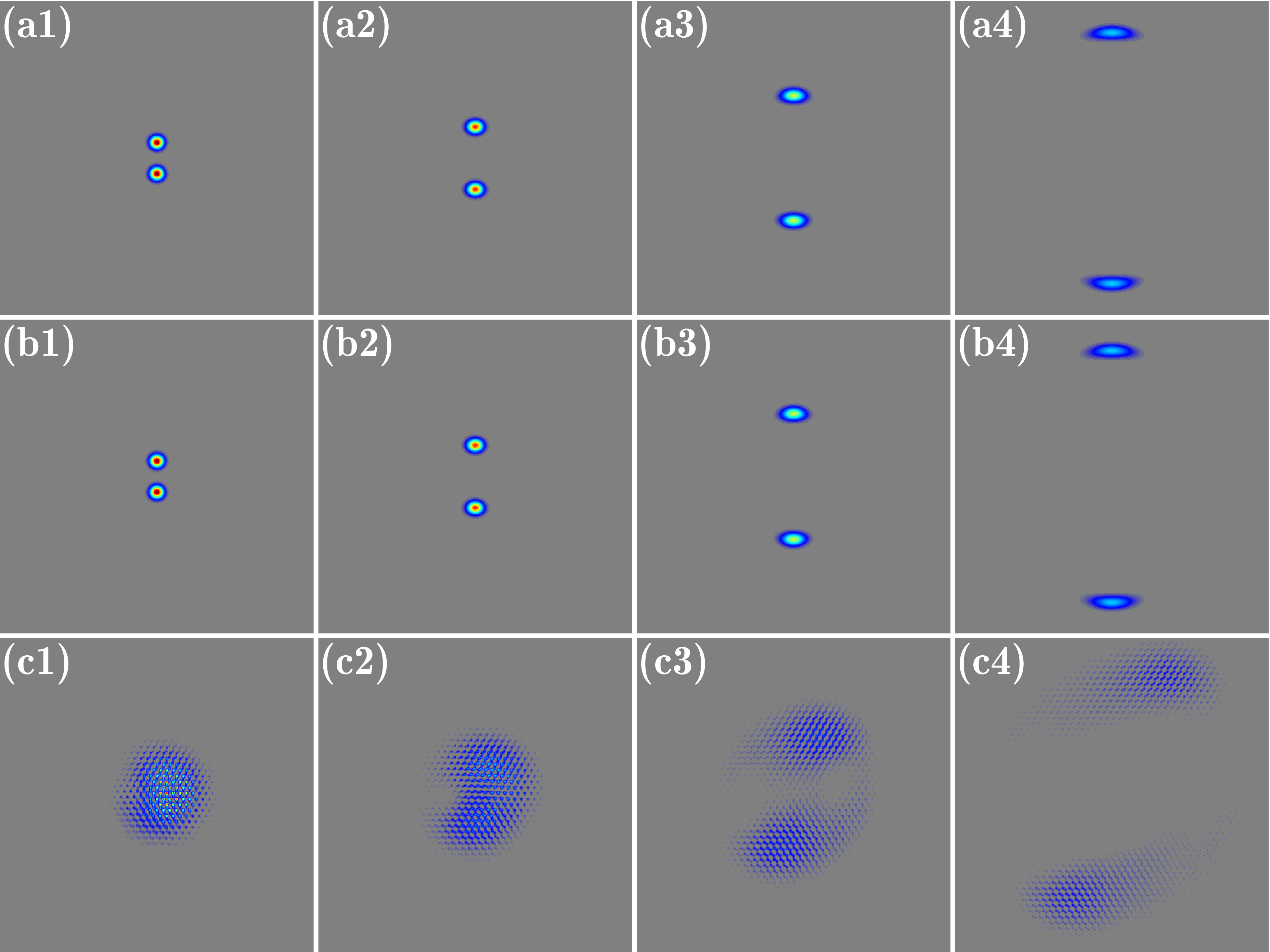}
\caption{Same as Fig. \ref{fig2}, but for obliquely incident light beams. }
\label{fig3}
\end{figure}

To further demonstrate the link between the FSE and the HCL,
we investigate the propagation of obliquely incident beams in the FSE, the DWE, and the HCL.
This is shown in Fig. \ref{fig3}, which is organized similar to Fig. \ref{fig2}.

In Figs. \ref{fig3}(a1)-\ref{fig3}(a4), the propagation of two obliquely incident Gaussian beams $\psi_{1,0}=\exp(-r^2/25)\exp(i2y)$
and $\psi_{2,0}=\exp(-r^2/25)\exp(-i2y)$ in the FSE is depicted.
According to the rule defined in Ref. \cite{zhang.sr.6.23645.2016}, the beam centers will be at $(x,y)=z(0,\pm1)$ during propagation;
that is, the two beams in propagation will separate linearly from each other in the plane $x=0$, as presented in Figs. \ref{fig3}(a1)-\ref{fig3}(a4).
Different from the FSE case, in the DWE case in Figs. \ref{fig3}(b1)-\ref{fig3}(b4),
we display the propagation of only one obliquely incident Gaussian beam $\psi_0=\exp(-r^2/25)\exp(i2y)$.
One finds that the obliquely incident Gaussian beam splits into two Gaussian beams during propagation,
and the behavior of these two Gaussian beams is the same as the behavior in Figs. \ref{fig3}(a1)-\ref{fig3}(a4).
For comparison, the propagation of a slightly oblique beam
that excites the Bloch mode of the Dirac cone is exhibited in Figs. \ref{fig3}(c1)-\ref{fig3}(c4).
One must remember that the dispersion can only be viewed as linear in a small region around the Dirac point,
and this is the reason why we choose a slightly oblique beam with a wide width.

The explanation of the propagation behavior observed in Fig. \ref{fig3} is quite direct.
As displayed in Fig. \ref{fig4}, we inspect the momentum spectra corresponding to FSE, DWE, and HCL.
In Fig. \ref{fig4}(a),  the input two Gaussian beams will excite the Bloch modes located at sites \textit{A} and \textit{B}, respectively.
Whereas in Fig. \ref{fig4}(b), only one input Gaussian beam can excite two Bloch modes at sites \textit{A} and \textit{B},
because the Bloch modes are degenerate only at the Dirac point, and the degeneracy will be lifted if there is a shift in the momentum space.
Since the dispersion along the polar direction is linear, as represented by the red lines in Figs. \ref{fig4}(a) and \ref{fig4}(b),
the beam width along the vertical direction in Figs. \ref{fig3}(a) and \ref{fig3}(b) does not change.
However, along the angular direction, as indicated by the red ellipses in Figs. \ref{fig4}(a) and \ref{fig4}(b), the dispersion is quadratic,
which means the beams will spread along the horizontal direction, as shown in Figs. \ref{fig3}(a) and \ref{fig3}(b).

\begin{figure}[htpb]
\centering
\includegraphics[width=0.45\columnwidth]{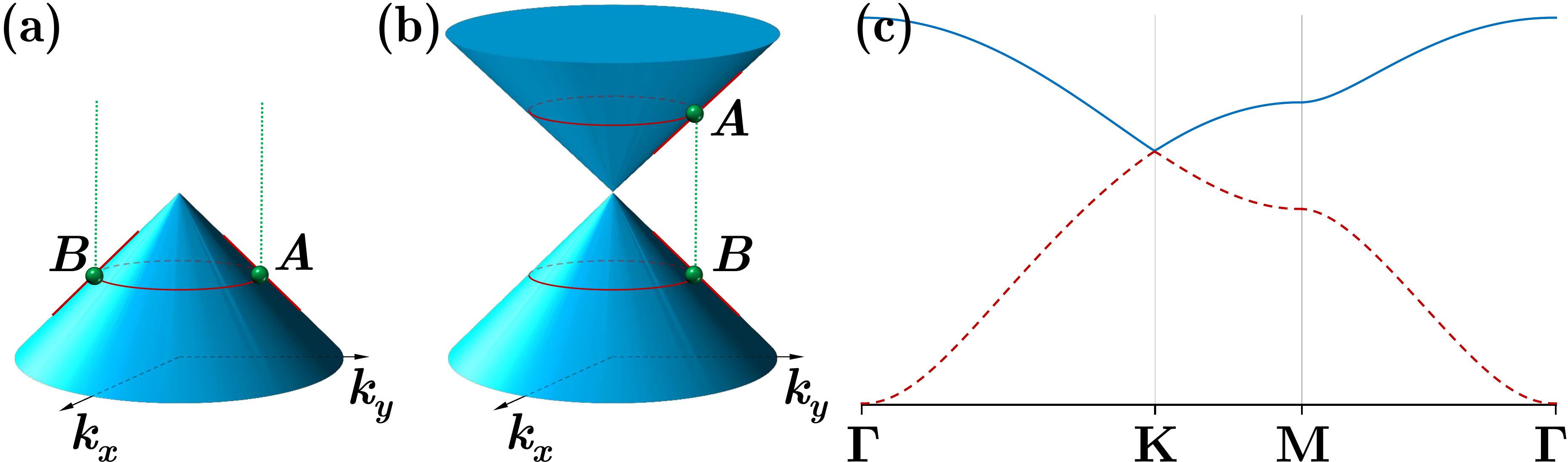}
\caption{Momentum spectra of (a) FSE, (b) DWE, and (c) HCL in the first Brillouin zone among high symmetric points.}
\label{fig4}
\end{figure}

One has reasons to believe that the two beams will evolve into a structure
that is close to the conical diffraction with the increasing propagation distance,
and the smaller the obliquity angle, the smaller the propagation distance to observe the conical-diffraction-like structure.
This is more readily observed in Fig. \ref{fig4}(c) than in Figs. \ref{fig4}(a) or \ref{fig4}(b).
In that figure, corresponding to Fig. \ref{fig1}(c), the band structure is shown along high-symmetry points.
The Dirac point is at \textbf{K}, and one cannot excite the Bloch mode far away from \textbf{K}, to guarantee the linear dispersion.
Therefore, one has to prepare the input beam that meets two conditions: large width and small slope.
In Fig. \ref{fig3}(c), the input beam is the three-wave interference pattern
which is multiplied by a wide Gaussian beam $\exp(-r^2/400)\exp(iy/10)$.
Since there are more than two bands in this continuum model, the wide input may excite additional Bloch modes belonging to other bands,
which will cause the movement of wide split beams in the circular direction.
Nevertheless, one still may agree that the phenomena observed in FSE can also be obtained by using HCL.
In other words, the HCL can well mimic the behavior in FSE.

One may raise the following question on this link: how if instead of free propagation,
an addition potential is considered in FSE? Will the HCL connection still be valid?
The answer is in the negative.
When, e.g., an additional harmonic potential is considered, Eq. (\ref{eq7}) will be modified as
\begin{equation}\label{eq10}
i \frac{\partial \psi}{\partial z} + \nabla^2\psi + V_h(x,y)\psi + V(x,y)\psi = 0,
\end{equation}
with $V(x,y)=-\sigma(x^2+y^2)/2$ and $\sigma$ determining the potential width.
Qualitatively, the harmonic potential is a large perturbation of the HCL,
so the properties of HCL will be affected greatly.
One dramatic change is that the translational symmetry of the HCL is violated,
which means that the parity symmetry of the HCL is broken.
Therefore, the Dirac cone will disappear \cite{lu.np.8.821.2014}, and the dispersion will never be linear.

In conclusion, we have investigated the relation between the FSE and the HCL,
and found that the connection can be established via the DWE.
Numerical simulations support our theoretical predictions.
However, such a link is not generally feasible, because when an additional potential is considered,
the parity symmetry of the HCL is broken, which leads to the disappearance of the Dirac cone.
We believe that our work is a significant attempt to find a real physical system that can be described by the FSE,
which may inspire new ideas on how to build a novel and dispersion-controllable optical system.

The work was supported by the National Basic Research Program of China (2012CB921804),
National Natural Science Foundation of China (61308015, 11474228,  61605154),
Key Scientific and Technological Innovation Team of Shaanxi Province (2014KCT-10),
and Qatar National Research Fund  (NPRP 6-021-1-005).
MRB acknowledges support by the Al Sraiya Holding Group.

\bibliography{my_refs_library}

\begin{thebibliography}{28}%
\makeatletter
\providecommand \@ifxundefined [1]{%
 \@ifx{#1\undefined}
}%
\providecommand \@ifnum [1]{%
 \ifnum #1\expandafter \@firstoftwo
 \else \expandafter \@secondoftwo
 \fi
}%
\providecommand \@ifx [1]{%
 \ifx #1\expandafter \@firstoftwo
 \else \expandafter \@secondoftwo
 \fi
}%
\providecommand \natexlab [1]{#1}%
\providecommand \enquote  [1]{``#1''}%
\providecommand \bibnamefont  [1]{#1}%
\providecommand \bibfnamefont [1]{#1}%
\providecommand \citenamefont [1]{#1}%
\providecommand \href@noop [0]{\@secondoftwo}%
\providecommand \href [0]{\begingroup \@sanitize@url \@href}%
\providecommand \@href[1]{\@@startlink{#1}\@@href}%
\providecommand \@@href[1]{\endgroup#1\@@endlink}%
\providecommand \@sanitize@url [0]{\catcode `\\12\catcode `\$12\catcode
  `\&12\catcode `\#12\catcode `\^12\catcode `\_12\catcode `\%12\relax}%
\providecommand \@@startlink[1]{}%
\providecommand \@@endlink[0]{}%
\providecommand \url  [0]{\begingroup\@sanitize@url \@url }%
\providecommand \@url [1]{\endgroup\@href {#1}{\urlprefix }}%
\providecommand \urlprefix  [0]{URL }%
\providecommand \Eprint [0]{\href }%
\providecommand \doibase [0]{http://dx.doi.org/}%
\providecommand \selectlanguage [0]{\@gobble}%
\providecommand \bibinfo  [0]{\@secondoftwo}%
\providecommand \bibfield  [0]{\@secondoftwo}%
\providecommand \translation [1]{[#1]}%
\providecommand \BibitemOpen [0]{}%
\providecommand \bibitemStop [0]{}%
\providecommand \bibitemNoStop [0]{.\EOS\space}%
\providecommand \EOS [0]{\spacefactor3000\relax}%
\providecommand \BibitemShut  [1]{\csname bibitem#1\endcsname}%
\let\auto@bib@innerbib\@empty
\bibitem [{\citenamefont
  {Laskin}(2000{\natexlab{a}})}]{laskin.pla.268.298.2000}%
  \BibitemOpen
  \bibfield  {author} {\bibinfo {author} {\bibfnamefont {N.}~\bibnamefont
  {Laskin}},\ }\emph {\bibinfo {title} {Fractional quantum mechanics and
  {L}\'{e}vy path integrals}},\ \href {\doibase 10.1016/S0375-9601(00)00201-2}
  {\bibfield  {journal} {\bibinfo  {journal} {Phys. Lett. A}\ }\textbf
  {\bibinfo {volume} {268}},\ \bibinfo {pages} {298} (\bibinfo {year}
  {2000}{\natexlab{a}})}\BibitemShut {NoStop}%
\bibitem [{\citenamefont
  {Laskin}(2000{\natexlab{b}})}]{laskin.pre.62.3135.2000}%
  \BibitemOpen
  \bibfield  {author} {\bibinfo {author} {\bibfnamefont {N.}~\bibnamefont
  {Laskin}},\ }\emph {\bibinfo {title} {Fractional quantum mechanics}},\ \href
  {\doibase 10.1103/PhysRevE.62.3135} {\bibfield  {journal} {\bibinfo
  {journal} {Phys. Rev. E}\ }\textbf {\bibinfo {volume} {62}},\ \bibinfo
  {pages} {3135} (\bibinfo {year} {2000}{\natexlab{b}})}\BibitemShut {NoStop}%
\bibitem [{\citenamefont {Laskin}(2002)}]{laskin.pre.66.056108.2002}%
  \BibitemOpen
  \bibfield  {author} {\bibinfo {author} {\bibfnamefont {N.}~\bibnamefont
  {Laskin}},\ }\emph {\bibinfo {title} {Fractional {S}chr\"odinger equation}},\
  \href {\doibase 10.1103/PhysRevE.66.056108} {\bibfield  {journal} {\bibinfo
  {journal} {Phys. Rev. E}\ }\textbf {\bibinfo {volume} {66}},\ \bibinfo
  {pages} {056108} (\bibinfo {year} {2002})}\BibitemShut {NoStop}%
\bibitem [{\citenamefont {Zhang}\ \emph
  {et~al.}(2015{\natexlab{a}})\citenamefont {Zhang}, \citenamefont {Liu},
  \citenamefont {Beli\'{c}}, \citenamefont {Zhong}, \citenamefont {Zhang},\
  and\ \citenamefont {Xiao}}]{zhang.prl.115.180403.2015}%
  \BibitemOpen
  \bibfield  {author} {\bibinfo {author} {\bibfnamefont {Y.~Q.}\ \bibnamefont
  {Zhang}}, \bibinfo {author} {\bibfnamefont {X.}~\bibnamefont {Liu}}, \bibinfo
  {author} {\bibfnamefont {M.~R.}\ \bibnamefont {Beli\'{c}}}, \bibinfo {author}
  {\bibfnamefont {W.~P.}\ \bibnamefont {Zhong}}, \bibinfo {author}
  {\bibfnamefont {Y.~P.}\ \bibnamefont {Zhang}}, \ and\ \bibinfo {author}
  {\bibfnamefont {M.}~\bibnamefont {Xiao}},\ }\emph {\bibinfo {title}
  {Propagation Dynamics of a Light Beam in a Fractional {S}chr\"odinger
  Equation}},\ \href {\doibase 10.1103/PhysRevLett.115.180403} {\bibfield
  {journal} {\bibinfo  {journal} {Phys. Rev. Lett.}\ }\textbf {\bibinfo
  {volume} {115}},\ \bibinfo {pages} {180403} (\bibinfo {year}
  {2015}{\natexlab{a}})}\BibitemShut {NoStop}%
\bibitem [{\citenamefont {Zhang}\ \emph
  {et~al.}(2016{\natexlab{a}})\citenamefont {Zhang}, \citenamefont {Zhong},
  \citenamefont {Beli\'c}, \citenamefont {Ahmed}, \citenamefont {Zhang},\ and\
  \citenamefont {Xiao}}]{zhang.sr.6.23645.2016}%
  \BibitemOpen
  \bibfield  {author} {\bibinfo {author} {\bibfnamefont {Y.~Q.}\ \bibnamefont
  {Zhang}}, \bibinfo {author} {\bibfnamefont {H.}~\bibnamefont {Zhong}},
  \bibinfo {author} {\bibfnamefont {M.~R.}\ \bibnamefont {Beli\'c}}, \bibinfo
  {author} {\bibfnamefont {N.}~\bibnamefont {Ahmed}}, \bibinfo {author}
  {\bibfnamefont {Y.~P.}\ \bibnamefont {Zhang}}, \ and\ \bibinfo {author}
  {\bibfnamefont {M.}~\bibnamefont {Xiao}},\ }\emph {\bibinfo {title}
  {Diffraction-free beams in fractional {S}chr\"odinger equation}},\ \href
  {\doibase 10.1038/srep23645} {\bibfield  {journal} {\bibinfo  {journal} {Sci.
  Rep.}\ }\textbf {\bibinfo {volume} {6}},\ \bibinfo {pages} {23645} (\bibinfo
  {year} {2016}{\natexlab{a}})}\BibitemShut {NoStop}%
\bibitem [{\citenamefont {Zhang}\ \emph
  {et~al.}(2016{\natexlab{b}})\citenamefont {Zhang}, \citenamefont {Zhong},
  \citenamefont {Beli\'c}, \citenamefont {Zhu}, \citenamefont {Zhong},
  \citenamefont {Zhang}, \citenamefont {Christodoulides},\ and\ \citenamefont
  {Xiao}}]{zhang.lpr.10.526.2016}%
  \BibitemOpen
  \bibfield  {author} {\bibinfo {author} {\bibfnamefont {Y.~Q.}\ \bibnamefont
  {Zhang}}, \bibinfo {author} {\bibfnamefont {H.}~\bibnamefont {Zhong}},
  \bibinfo {author} {\bibfnamefont {M.~R.}\ \bibnamefont {Beli\'c}}, \bibinfo
  {author} {\bibfnamefont {Y.}~\bibnamefont {Zhu}}, \bibinfo {author}
  {\bibfnamefont {W.~P.}\ \bibnamefont {Zhong}}, \bibinfo {author}
  {\bibfnamefont {Y.~P.}\ \bibnamefont {Zhang}}, \bibinfo {author}
  {\bibfnamefont {D.~N.}\ \bibnamefont {Christodoulides}}, \ and\ \bibinfo
  {author} {\bibfnamefont {M.}~\bibnamefont {Xiao}},\ }\emph {\bibinfo {title}
  {PT symmetry in a fractional {S}chr\"odinger equation}},\ \href {\doibase
  10.1002/lpor.201600037} {\bibfield  {journal} {\bibinfo  {journal} {Laser
  Photon. Rev.}\ }\textbf {\bibinfo {volume} {10}},\ \bibinfo {pages} {526}
  (\bibinfo {year} {2016}{\natexlab{b}})}\BibitemShut {NoStop}%
\bibitem [{\citenamefont {Liemert}\ and\ \citenamefont
  {Kienle}(2016)}]{liemert.math.4.31.2016}%
  \BibitemOpen
  \bibfield  {author} {\bibinfo {author} {\bibfnamefont {A.}~\bibnamefont
  {Liemert}}\ and\ \bibinfo {author} {\bibfnamefont {A.}~\bibnamefont
  {Kienle}},\ }\emph {\bibinfo {title} {Fractional {S}chr\"odinger Equation in
  the Presence of the Linear Potential}},\ \href {\doibase 10.3390/math4020031}
  {\bibfield  {journal} {\bibinfo  {journal} {Mathematics}\ }\textbf {\bibinfo
  {volume} {4}},\ \bibinfo {pages} {31} (\bibinfo {year} {2016})}\BibitemShut
  {NoStop}%
\bibitem [{\citenamefont {Guo}\ and\ \citenamefont
  {Huang}(2012)}]{guo.jmp.53.083702.2012}%
  \BibitemOpen
  \bibfield  {author} {\bibinfo {author} {\bibfnamefont {B.}~\bibnamefont
  {Guo}}\ and\ \bibinfo {author} {\bibfnamefont {D.}~\bibnamefont {Huang}},\
  }\emph {\bibinfo {title} {Existence and stability of standing waves for
  nonlinear fractional {S}chr\"odinger equations}},\ \href {\doibase
  10.1063/1.4746806} {\bibfield  {journal} {\bibinfo  {journal} {J. Math.
  Phys.}\ }\textbf {\bibinfo {volume} {53}},\ \bibinfo {pages} {083702}
  (\bibinfo {year} {2012})}\BibitemShut {NoStop}%
\bibitem [{\citenamefont {Klein}\ \emph {et~al.}(2014)\citenamefont {Klein},
  \citenamefont {Sparber},\ and\ \citenamefont
  {Markowich}}]{klein.prsa.470.20140364.2014}%
  \BibitemOpen
  \bibfield  {author} {\bibinfo {author} {\bibfnamefont {C.}~\bibnamefont
  {Klein}}, \bibinfo {author} {\bibfnamefont {C.}~\bibnamefont {Sparber}}, \
  and\ \bibinfo {author} {\bibfnamefont {P.}~\bibnamefont {Markowich}},\ }\emph
  {\bibinfo {title} {Numerical study of fractional nonlinear {S}chr\"odinger
  equations}},\ \href {\doibase 10.1098/rspa.2014.0364} {\bibfield  {journal}
  {\bibinfo  {journal} {Proc. R.Soc. A}\ }\textbf {\bibinfo {volume} {470}},\
  \bibinfo {pages} {20140364} (\bibinfo {year} {2014})}\BibitemShut {NoStop}%
\bibitem [{\citenamefont {Zhang}\ \emph
  {et~al.}(2016{\natexlab{c}})\citenamefont {Zhang}, \citenamefont {Li},
  \citenamefont {Zhong}, \citenamefont {Xu}, \citenamefont {Lei}, \citenamefont
  {Li},\ and\ \citenamefont {Fan}}]{zhang.oe.24.14406.2016}%
  \BibitemOpen
  \bibfield  {author} {\bibinfo {author} {\bibfnamefont {L.}~\bibnamefont
  {Zhang}}, \bibinfo {author} {\bibfnamefont {C.}~\bibnamefont {Li}}, \bibinfo
  {author} {\bibfnamefont {H.}~\bibnamefont {Zhong}}, \bibinfo {author}
  {\bibfnamefont {C.}~\bibnamefont {Xu}}, \bibinfo {author} {\bibfnamefont
  {D.}~\bibnamefont {Lei}}, \bibinfo {author} {\bibfnamefont {Y.}~\bibnamefont
  {Li}}, \ and\ \bibinfo {author} {\bibfnamefont {D.}~\bibnamefont {Fan}},\
  }\emph {\bibinfo {title} {Propagation dynamics of super-{G}aussian beams in
  fractional {S}chr\"odinger equation: from linear to nonlinear regimes}},\
  \href {\doibase 10.1364/OE.24.014406} {\bibfield  {journal} {\bibinfo
  {journal} {Opt. Express}\ }\textbf {\bibinfo {volume} {24}},\ \bibinfo
  {pages} {14406} (\bibinfo {year} {2016}{\natexlab{c}})}\BibitemShut {NoStop}%
\bibitem [{\citenamefont {Longhi}(2015)}]{longhi.ol.36.2883.2015}%
  \BibitemOpen
  \bibfield  {author} {\bibinfo {author} {\bibfnamefont {S.}~\bibnamefont
  {Longhi}},\ }\emph {\bibinfo {title} {Fractional {S}chr\"odinger equation in
  optics}},\ \href {\doibase 10.1364/OL.40.001117} {\bibfield  {journal}
  {\bibinfo  {journal} {Opt. Lett.}\ }\textbf {\bibinfo {volume} {40}},\
  \bibinfo {pages} {1117} (\bibinfo {year} {2015})}\BibitemShut {NoStop}%
\bibitem [{\citenamefont {Lu}\ \emph {et~al.}(2014)\citenamefont {Lu},
  \citenamefont {Joannopoulos},\ and\ \citenamefont
  {Solja{\v{c}}i{\'c}}}]{lu.np.8.821.2014}%
  \BibitemOpen
  \bibfield  {author} {\bibinfo {author} {\bibfnamefont {L.}~\bibnamefont
  {Lu}}, \bibinfo {author} {\bibfnamefont {J.~D.}\ \bibnamefont
  {Joannopoulos}}, \ and\ \bibinfo {author} {\bibfnamefont {M.}~\bibnamefont
  {Solja{\v{c}}i{\'c}}},\ }\emph {\bibinfo {title} {Topological photonics}},\
  \href {\doibase 10.1038/nphoton.2014.248} {\bibfield  {journal} {\bibinfo
  {journal} {Nat. Photon.}\ }\textbf {\bibinfo {volume} {8}},\ \bibinfo {pages}
  {821} (\bibinfo {year} {2014})}\BibitemShut {NoStop}%
\bibitem [{\citenamefont {Rechtsman}\ \emph {et~al.}(2013)\citenamefont
  {Rechtsman}, \citenamefont {Zeuner}, \citenamefont {Plotnik}, \citenamefont
  {Lumer}, \citenamefont {Podolsky}, \citenamefont {Dreisow}, \citenamefont
  {Nolte}, \citenamefont {Segev},\ and\ \citenamefont
  {Szameit}}]{rechtsman.nature.496.196.2013}%
  \BibitemOpen
  \bibfield  {author} {\bibinfo {author} {\bibfnamefont {M.~C.}\ \bibnamefont
  {Rechtsman}}, \bibinfo {author} {\bibfnamefont {J.~M.}\ \bibnamefont
  {Zeuner}}, \bibinfo {author} {\bibfnamefont {Y.}~\bibnamefont {Plotnik}},
  \bibinfo {author} {\bibfnamefont {Y.}~\bibnamefont {Lumer}}, \bibinfo
  {author} {\bibfnamefont {D.}~\bibnamefont {Podolsky}}, \bibinfo {author}
  {\bibfnamefont {F.}~\bibnamefont {Dreisow}}, \bibinfo {author} {\bibfnamefont
  {S.}~\bibnamefont {Nolte}}, \bibinfo {author} {\bibfnamefont
  {M.}~\bibnamefont {Segev}}, \ and\ \bibinfo {author} {\bibfnamefont
  {A.}~\bibnamefont {Szameit}},\ }\emph {\bibinfo {title} {Photonic {F}loquet
  topological insulators}},\ \href {\doibase 10.1038/nature12066} {\bibfield
  {journal} {\bibinfo  {journal} {Nature}\ }\textbf {\bibinfo {volume} {496}},\
  \bibinfo {pages} {196} (\bibinfo {year} {2013})}\BibitemShut {NoStop}%
\bibitem [{\citenamefont {Plotnik}\ \emph {et~al.}(2014)\citenamefont
  {Plotnik}, \citenamefont {Rechtsman}, \citenamefont {Song}, \citenamefont
  {Heinrich}, \citenamefont {Zeuner}, \citenamefont {Nolte}, \citenamefont
  {Lumer}, \citenamefont {Malkova}, \citenamefont {Xu}, \citenamefont
  {Szameit}, \citenamefont {Chen},\ and\ \citenamefont
  {Segev}}]{plotnik.nm.13.57.2014}%
  \BibitemOpen
  \bibfield  {author} {\bibinfo {author} {\bibfnamefont {Y.}~\bibnamefont
  {Plotnik}}, \bibinfo {author} {\bibfnamefont {M.~C.}\ \bibnamefont
  {Rechtsman}}, \bibinfo {author} {\bibfnamefont {D.}~\bibnamefont {Song}},
  \bibinfo {author} {\bibfnamefont {M.}~\bibnamefont {Heinrich}}, \bibinfo
  {author} {\bibfnamefont {J.~M.}\ \bibnamefont {Zeuner}}, \bibinfo {author}
  {\bibfnamefont {S.}~\bibnamefont {Nolte}}, \bibinfo {author} {\bibfnamefont
  {Y.}~\bibnamefont {Lumer}}, \bibinfo {author} {\bibfnamefont
  {N.}~\bibnamefont {Malkova}}, \bibinfo {author} {\bibfnamefont
  {J.}~\bibnamefont {Xu}}, \bibinfo {author} {\bibfnamefont {A.}~\bibnamefont
  {Szameit}}, \bibinfo {author} {\bibfnamefont {Z.}~\bibnamefont {Chen}}, \
  and\ \bibinfo {author} {\bibfnamefont {M.}~\bibnamefont {Segev}},\ }\emph
  {\bibinfo {title} {Observation of unconventional edge states in `photonic
  graphene'}},\ \href {\doibase 10.1038/nmat3783} {\bibfield  {journal}
  {\bibinfo  {journal} {Nat. Mater.}\ }\textbf {\bibinfo {volume} {13}},\
  \bibinfo {pages} {57} (\bibinfo {year} {2014})}\BibitemShut {NoStop}%
\bibitem [{\citenamefont {Castro~Neto}\ \emph {et~al.}(2009)\citenamefont
  {Castro~Neto}, \citenamefont {Guinea}, \citenamefont {Peres}, \citenamefont
  {Novoselov},\ and\ \citenamefont {Geim}}]{neto.rmp.81.109.2009}%
  \BibitemOpen
  \bibfield  {author} {\bibinfo {author} {\bibfnamefont {A.~H.}\ \bibnamefont
  {Castro~Neto}}, \bibinfo {author} {\bibfnamefont {F.}~\bibnamefont {Guinea}},
  \bibinfo {author} {\bibfnamefont {N.~M.~R.}\ \bibnamefont {Peres}}, \bibinfo
  {author} {\bibfnamefont {K.~S.}\ \bibnamefont {Novoselov}}, \ and\ \bibinfo
  {author} {\bibfnamefont {A.~K.}\ \bibnamefont {Geim}},\ }\emph {\bibinfo
  {title} {The electronic properties of graphene}},\ \href {\doibase
  10.1103/RevModPhys.81.109} {\bibfield  {journal} {\bibinfo  {journal} {Rev.
  Mod. Phys.}\ }\textbf {\bibinfo {volume} {81}},\ \bibinfo {pages} {109}
  (\bibinfo {year} {2009})}\BibitemShut {NoStop}%
\bibitem [{\citenamefont {Ablowitz}\ \emph {et~al.}(2009)\citenamefont
  {Ablowitz}, \citenamefont {Nixon},\ and\ \citenamefont
  {Zhu}}]{ablowitz.pra.79.053830.2009}%
  \BibitemOpen
  \bibfield  {author} {\bibinfo {author} {\bibfnamefont {M.~J.}\ \bibnamefont
  {Ablowitz}}, \bibinfo {author} {\bibfnamefont {S.~D.}\ \bibnamefont {Nixon}},
  \ and\ \bibinfo {author} {\bibfnamefont {Y.}~\bibnamefont {Zhu}},\ }\emph
  {\bibinfo {title} {Conical diffraction in honeycomb lattices}},\ \href
  {\doibase 10.1103/PhysRevA.79.053830} {\bibfield  {journal} {\bibinfo
  {journal} {Phys. Rev. A}\ }\textbf {\bibinfo {volume} {79}},\ \bibinfo
  {pages} {053830} (\bibinfo {year} {2009})}\BibitemShut {NoStop}%
\bibitem [{\citenamefont {Ablowitz}\ and\ \citenamefont
  {Zhu}(2010)}]{ablowitz.pra.82.013840.2010}%
  \BibitemOpen
  \bibfield  {author} {\bibinfo {author} {\bibfnamefont {M.~J.}\ \bibnamefont
  {Ablowitz}}\ and\ \bibinfo {author} {\bibfnamefont {Y.}~\bibnamefont {Zhu}},\
  }\emph {\bibinfo {title} {Evolution of {B}loch-mode envelopes in
  two-dimensional generalized honeycomb lattices}},\ \href {\doibase
  10.1103/PhysRevA.82.013840} {\bibfield  {journal} {\bibinfo  {journal} {Phys.
  Rev. A}\ }\textbf {\bibinfo {volume} {82}},\ \bibinfo {pages} {013840}
  (\bibinfo {year} {2010})}\BibitemShut {NoStop}%
\bibitem [{\citenamefont {Song}\ \emph
  {et~al.}(2015{\natexlab{a}})\citenamefont {Song}, \citenamefont {Paltoglou},
  \citenamefont {Liu}, \citenamefont {Zhu}, \citenamefont {Gallardo},
  \citenamefont {Tang}, \citenamefont {Xu}, \citenamefont {Ablowitz},
  \citenamefont {Efremidis},\ and\ \citenamefont {Chen}}]{song.nc.6.6272.2015}%
  \BibitemOpen
  \bibfield  {author} {\bibinfo {author} {\bibfnamefont {D.}~\bibnamefont
  {Song}}, \bibinfo {author} {\bibfnamefont {V.}~\bibnamefont {Paltoglou}},
  \bibinfo {author} {\bibfnamefont {S.}~\bibnamefont {Liu}}, \bibinfo {author}
  {\bibfnamefont {Y.}~\bibnamefont {Zhu}}, \bibinfo {author} {\bibfnamefont
  {D.}~\bibnamefont {Gallardo}}, \bibinfo {author} {\bibfnamefont
  {L.}~\bibnamefont {Tang}}, \bibinfo {author} {\bibfnamefont {J.}~\bibnamefont
  {Xu}}, \bibinfo {author} {\bibfnamefont {M.}~\bibnamefont {Ablowitz}},
  \bibinfo {author} {\bibfnamefont {N.~K.}\ \bibnamefont {Efremidis}}, \ and\
  \bibinfo {author} {\bibfnamefont {Z.}~\bibnamefont {Chen}},\ }\emph {\bibinfo
  {title} {Unveiling pseudospin and angular momentum in photonic graphene.}},\
  \href {\doibase 10.1038/ncomms7272} {\bibfield  {journal} {\bibinfo
  {journal} {Nat. Commun.}\ }\textbf {\bibinfo {volume} {6}},\ \bibinfo {pages}
  {6272} (\bibinfo {year} {2015}{\natexlab{a}})}\BibitemShut {NoStop}%
\bibitem [{\citenamefont {Song}\ \emph
  {et~al.}(2015{\natexlab{b}})\citenamefont {Song}, \citenamefont {Liu},
  \citenamefont {Paltoglou}, \citenamefont {Gallardo}, \citenamefont {Tang},
  \citenamefont {Zhao}, \citenamefont {Xu}, \citenamefont {Efremidis},\ and\
  \citenamefont {Chen}}]{song.2dm.2.034007.2015}%
  \BibitemOpen
  \bibfield  {author} {\bibinfo {author} {\bibfnamefont {D.}~\bibnamefont
  {Song}}, \bibinfo {author} {\bibfnamefont {S.}~\bibnamefont {Liu}}, \bibinfo
  {author} {\bibfnamefont {V.}~\bibnamefont {Paltoglou}}, \bibinfo {author}
  {\bibfnamefont {D.}~\bibnamefont {Gallardo}}, \bibinfo {author}
  {\bibfnamefont {L.}~\bibnamefont {Tang}}, \bibinfo {author} {\bibfnamefont
  {J.}~\bibnamefont {Zhao}}, \bibinfo {author} {\bibfnamefont {J.}~\bibnamefont
  {Xu}}, \bibinfo {author} {\bibfnamefont {N.~K.}\ \bibnamefont {Efremidis}}, \
  and\ \bibinfo {author} {\bibfnamefont {Z.}~\bibnamefont {Chen}},\ }\emph
  {\bibinfo {title} {Controlled generation of pseudospin-mediated vortices in
  photonic graphene}},\ \href {http://stacks.iop.org/2053-1583/2/i=3/a=034007}
  {\bibfield  {journal} {\bibinfo  {journal} {2D Mater.}\ }\textbf {\bibinfo
  {volume} {2}},\ \bibinfo {pages} {034007} (\bibinfo {year}
  {2015}{\natexlab{b}})}\BibitemShut {NoStop}%
\bibitem [{\citenamefont {Ablowitz}\ and\ \citenamefont
  {Zhu}(2012)}]{ablowitz.siam.72.240.2012}%
  \BibitemOpen
  \bibfield  {author} {\bibinfo {author} {\bibfnamefont {M.~J.}\ \bibnamefont
  {Ablowitz}}\ and\ \bibinfo {author} {\bibfnamefont {Y.}~\bibnamefont {Zhu}},\
  }\emph {\bibinfo {title} {Nonlinear waves in shallow honeycomb lattices}},\
  \href {\doibase 10.1137/11082662X} {\bibfield  {journal} {\bibinfo  {journal}
  {SIAM J. Appl. Math.}\ }\textbf {\bibinfo {volume} {72}},\ \bibinfo {pages}
  {240} (\bibinfo {year} {2012})}\BibitemShut {NoStop}%
\bibitem [{\citenamefont {Boguslawski}\ \emph {et~al.}(2011)\citenamefont
  {Boguslawski}, \citenamefont {Rose},\ and\ \citenamefont
  {Denz}}]{boguslawski.pra.84.013832.2011}%
  \BibitemOpen
  \bibfield  {author} {\bibinfo {author} {\bibfnamefont {M.}~\bibnamefont
  {Boguslawski}}, \bibinfo {author} {\bibfnamefont {P.}~\bibnamefont {Rose}}, \
  and\ \bibinfo {author} {\bibfnamefont {C.}~\bibnamefont {Denz}},\ }\emph
  {\bibinfo {title} {Increasing the structural variety of discrete
  nondiffracting wave fields}},\ \href {\doibase 10.1103/PhysRevA.84.013832}
  {\bibfield  {journal} {\bibinfo  {journal} {Phys. Rev. A}\ }\textbf {\bibinfo
  {volume} {84}},\ \bibinfo {pages} {013832} (\bibinfo {year}
  {2011})}\BibitemShut {NoStop}%
\bibitem [{\citenamefont {Bartal}\ \emph {et~al.}(2005)\citenamefont {Bartal},
  \citenamefont {Cohen}, \citenamefont {Buljan}, \citenamefont {Fleischer},
  \citenamefont {Manela},\ and\ \citenamefont
  {Segev}}]{bartal.prl.94.163902.2005}%
  \BibitemOpen
  \bibfield  {author} {\bibinfo {author} {\bibfnamefont {G.}~\bibnamefont
  {Bartal}}, \bibinfo {author} {\bibfnamefont {O.}~\bibnamefont {Cohen}},
  \bibinfo {author} {\bibfnamefont {H.}~\bibnamefont {Buljan}}, \bibinfo
  {author} {\bibfnamefont {J.~W.}\ \bibnamefont {Fleischer}}, \bibinfo {author}
  {\bibfnamefont {O.}~\bibnamefont {Manela}}, \ and\ \bibinfo {author}
  {\bibfnamefont {M.}~\bibnamefont {Segev}},\ }\emph {\bibinfo {title}
  {Brillouin Zone Spectroscopy of Nonlinear Photonic Lattices}},\ \href
  {\doibase 10.1103/PhysRevLett.94.163902} {\bibfield  {journal} {\bibinfo
  {journal} {Phys. Rev. Lett.}\ }\textbf {\bibinfo {volume} {94}},\ \bibinfo
  {pages} {163902} (\bibinfo {year} {2005})}\BibitemShut {NoStop}%
\bibitem [{\citenamefont {Freedman}\ \emph {et~al.}(2006)\citenamefont
  {Freedman}, \citenamefont {Bartal}, \citenamefont {Segev}, \citenamefont
  {Lifshitz}, \citenamefont {Christodoulides},\ and\ \citenamefont
  {Fleischer}}]{freedman.nature.440.1166.2006}%
  \BibitemOpen
  \bibfield  {author} {\bibinfo {author} {\bibfnamefont {B.}~\bibnamefont
  {Freedman}}, \bibinfo {author} {\bibfnamefont {G.}~\bibnamefont {Bartal}},
  \bibinfo {author} {\bibfnamefont {M.}~\bibnamefont {Segev}}, \bibinfo
  {author} {\bibfnamefont {R.}~\bibnamefont {Lifshitz}}, \bibinfo {author}
  {\bibfnamefont {D.~N.}\ \bibnamefont {Christodoulides}}, \ and\ \bibinfo
  {author} {\bibfnamefont {J.~W.}\ \bibnamefont {Fleischer}},\ }\emph {\bibinfo
  {title} {Wave and defect dynamics in nonlinear photonic quasicrystals}},\
  \href {\doibase 10.1038/nature04722} {\bibfield  {journal} {\bibinfo
  {journal} {Nature}\ }\textbf {\bibinfo {volume} {440}},\ \bibinfo {pages}
  {1166} (\bibinfo {year} {2006})}\BibitemShut {NoStop}%
\bibitem [{\citenamefont {Liu}\ \emph {et~al.}(2010)\citenamefont {Liu},
  \citenamefont {Zhang}, \citenamefont {Gan}, \citenamefont {Xiao},\ and\
  \citenamefont {Zhao}}]{liu.apb.99.727.2010}%
  \BibitemOpen
  \bibfield  {author} {\bibinfo {author} {\bibfnamefont {S.}~\bibnamefont
  {Liu}}, \bibinfo {author} {\bibfnamefont {P.}~\bibnamefont {Zhang}}, \bibinfo
  {author} {\bibfnamefont {X.}~\bibnamefont {Gan}}, \bibinfo {author}
  {\bibfnamefont {F.}~\bibnamefont {Xiao}}, \ and\ \bibinfo {author}
  {\bibfnamefont {J.}~\bibnamefont {Zhao}},\ }\emph {\bibinfo {title}
  {Visualization of the {B}ragg reflection in complex photonic lattices by
  employing {B}rillouin zone spectroscopy}},\ \href {\doibase
  10.1007/s00340-010-4042-6} {\bibfield  {journal} {\bibinfo  {journal} {Appl.
  Phys. B}\ }\textbf {\bibinfo {volume} {99}},\ \bibinfo {pages} {727}
  (\bibinfo {year} {2010})}\BibitemShut {NoStop}%
\bibitem [{\citenamefont {Zhang}\ \emph {et~al.}(2010)\citenamefont {Zhang},
  \citenamefont {Liu}, \citenamefont {Lou}, \citenamefont {Xiao}, \citenamefont
  {Wang}, \citenamefont {Zhao}, \citenamefont {Xu},\ and\ \citenamefont
  {Chen}}]{zhang.pra.81.041801.2010}%
  \BibitemOpen
  \bibfield  {author} {\bibinfo {author} {\bibfnamefont {P.}~\bibnamefont
  {Zhang}}, \bibinfo {author} {\bibfnamefont {S.}~\bibnamefont {Liu}}, \bibinfo
  {author} {\bibfnamefont {C.}~\bibnamefont {Lou}}, \bibinfo {author}
  {\bibfnamefont {F.}~\bibnamefont {Xiao}}, \bibinfo {author} {\bibfnamefont
  {X.}~\bibnamefont {Wang}}, \bibinfo {author} {\bibfnamefont {J.}~\bibnamefont
  {Zhao}}, \bibinfo {author} {\bibfnamefont {J.}~\bibnamefont {Xu}}, \ and\
  \bibinfo {author} {\bibfnamefont {Z.}~\bibnamefont {Chen}},\ }\emph {\bibinfo
  {title} {Incomplete {B}rillouin-zone spectra and controlled {B}ragg
  reflection with ionic-type photonic lattices}},\ \href {\doibase
  10.1103/PhysRevA.81.041801} {\bibfield  {journal} {\bibinfo  {journal} {Phys.
  Rev. A}\ }\textbf {\bibinfo {volume} {81}},\ \bibinfo {pages} {041801}
  (\bibinfo {year} {2010})}\BibitemShut {NoStop}%
\bibitem [{\citenamefont {Peleg}\ \emph {et~al.}(2007)\citenamefont {Peleg},
  \citenamefont {Bartal}, \citenamefont {Freedman}, \citenamefont {Manela},
  \citenamefont {Segev},\ and\ \citenamefont
  {Christodoulides}}]{peleg.prl.98.103901.2007}%
  \BibitemOpen
  \bibfield  {author} {\bibinfo {author} {\bibfnamefont {O.}~\bibnamefont
  {Peleg}}, \bibinfo {author} {\bibfnamefont {G.}~\bibnamefont {Bartal}},
  \bibinfo {author} {\bibfnamefont {B.}~\bibnamefont {Freedman}}, \bibinfo
  {author} {\bibfnamefont {O.}~\bibnamefont {Manela}}, \bibinfo {author}
  {\bibfnamefont {M.}~\bibnamefont {Segev}}, \ and\ \bibinfo {author}
  {\bibfnamefont {D.~N.}\ \bibnamefont {Christodoulides}},\ }\emph {\bibinfo
  {title} {Conical Diffraction and Gap Solitons in Honeycomb Photonic
  Lattices}},\ \href {\doibase 10.1103/PhysRevLett.98.103901} {\bibfield
  {journal} {\bibinfo  {journal} {Phys. Rev. Lett.}\ }\textbf {\bibinfo
  {volume} {98}},\ \bibinfo {pages} {103901} (\bibinfo {year}
  {2007})}\BibitemShut {NoStop}%
\bibitem [{\citenamefont {Zhang}\ \emph
  {et~al.}(2015{\natexlab{b}})\citenamefont {Zhang}, \citenamefont {Wu},
  \citenamefont {Beli\'c}, \citenamefont {Zheng}, \citenamefont {Wang},
  \citenamefont {Xiao},\ and\ \citenamefont {Zhang}}]{zhang.lpr.9.331.2015}%
  \BibitemOpen
  \bibfield  {author} {\bibinfo {author} {\bibfnamefont {Y.~Q.}\ \bibnamefont
  {Zhang}}, \bibinfo {author} {\bibfnamefont {Z.~K.}\ \bibnamefont {Wu}},
  \bibinfo {author} {\bibfnamefont {M.~R.}\ \bibnamefont {Beli\'c}}, \bibinfo
  {author} {\bibfnamefont {H.~B.}\ \bibnamefont {Zheng}}, \bibinfo {author}
  {\bibfnamefont {Z.~G.}\ \bibnamefont {Wang}}, \bibinfo {author}
  {\bibfnamefont {M.}~\bibnamefont {Xiao}}, \ and\ \bibinfo {author}
  {\bibfnamefont {Y.~P.}\ \bibnamefont {Zhang}},\ }\emph {\bibinfo {title}
  {Photonic {F}loquet topological insulators in atomic ensembles}},\ \href
  {\doibase 10.1002/lpor.201400428} {\bibfield  {journal} {\bibinfo  {journal}
  {Laser Photon. Rev.}\ }\textbf {\bibinfo {volume} {9}},\ \bibinfo {pages}
  {331} (\bibinfo {year} {2015}{\natexlab{b}})}\BibitemShut {NoStop}%
\bibitem [{\citenamefont {Berry}\ and\ \citenamefont
  {Jeffrey}(2007)}]{berry.po.50.13.2007}%
  \BibitemOpen
  \bibfield  {author} {\bibinfo {author} {\bibfnamefont {M.}~\bibnamefont
  {Berry}}\ and\ \bibinfo {author} {\bibfnamefont {M.}~\bibnamefont
  {Jeffrey}},\ }\emph {\bibinfo {title} {Conical diffraction: {H}amilton's
  diabolical point at the heart of crystal optics}},\ \href {\doibase
  10.1016/S0079-6638(07)50002-8} {\bibfield  {journal} {\bibinfo  {journal}
  {Prog. Opt.}\ }\textbf {\bibinfo {volume} {50}},\ \bibinfo {pages} {13}
  (\bibinfo {year} {2007})}\BibitemShut {NoStop}%
\end{thebibliography}%
\bibliographystyle{myprx}

\end{document}